\documentclass{optica-article}

\journal{opticajournal} 

\articletype{Research Article}

\usepackage{lineno}

\usepackage{graphicx}

\graphicspath{{picture/}}

\usepackage{xcolor}

\newcommand{\fst}[1]{#1}

\begin{document}

\title{Surface figure metrology for reflective membrane mirrors based on phase-measuring deflectometry}

\author{
Xin Yan,\authormark{1,2,3} 
Zhi-Kang Zhuang,\authormark{2,3} 
Fu-Jia Du,\authormark{2,3}
Wen Duan,\authormark{2,3}
Peilin Yin,\authormark{1}
Mo-Nong Yu,\authormark{4}
and
Guang Yang,\authormark{2,3,*}
}

\address{
\authormark{1}College of Science, Nanjing University of Posts and Telecommunications, Nanjing 210023, China\\
\authormark{2}Nanjing Institute of Astronomical Optics \& Technology, Chinese Academy of Sciences, Nanjing 210042, China\\
\authormark{3}CAS Key Laboratory of Astronomical Optics \& Technology, Nanjing Institute of Astronomical Optics \& Technology, Nanjing 210042, China\\
\authormark{4}Purple Mountain Observatory, Chinese Academy of Sciences, Nanjing 210023, China
}

\email{\authormark{*}gyang@niaot.ac.cn} 


\begin{abstract*} 
Reflective membrane mirrors provide a lightweight, low-cost alternative to traditional optics for next-generation large-aperture telescopes, but their non-rigid, thin structure poses challenges for surface metrology. 
We present a phase-measuring deflectometry (PMD) system enhanced with tailored ray-tracing and iterative reconstruction to enable non-contact measurement of large membrane optics.
The system successfully characterizes the surface figure and evaluates the dynamic stability of a 1-meter Hencky-type membrane mirror. Our results demonstrate the effectiveness of PMD as a practical metrology tool for future membrane-based telescope systems.

\end{abstract*}

\section{Introduction}

The light-gathering power of astronomical telescopes scales fundamentally with aperture size. From Galileo's centimeter-scale lenses in the early 17th century \cite{VanHelden10} to nowadays 39.3-meter Extremely Large Telescope (ELT) and 6.5-meter James Webb Space Telescope (JWST), four centuries of progress have been driven by aperture growth. Yet scaling beyond current sizes poses formidable challenges.
For space telescopes, launch mass constraints prohibit traditional glass/metal/ceramic mirrors ($\gtrsim 10$~kg~m$^{-2}$ areal density; e.g., \cite{stahl24}) from exceeding $\sim 10$-meter apertures. Ground-based telescopes face analogous cost barriers, where mirror weight compounds structural complexity for support and pointing systems \cite{marchiori2018, colussi2020}.

Membrane mirrors, thin ($\lesssim 100\ \mu$m) polymer sheets (e.g., Kapton and Mylar) with reflective metal coatings, offer a breakthrough with areal densities $\lesssim 0.1$~kg~m$^{-2}$. This enables proposed giant space observatories such as Orbiting Astronomical Satellite for Investigating Stellar Systems (OASIS, 14~m \cite{arenberg21}), Single Aperture Large Telescope for Universe Studies (SALTUS, 14~m \cite{harding24}), and even Kilometer Space Telescope (KST, $\gtrsim 1000$~m \cite{johnson21}). However, their compliant nature demands novel metrology approaches, as conventional contact methods such as laser trackers and coordinate-measuring machines risk distorting the surface \fst{\cite{wang20}}, while interferometers lack the required dynamic range for millimeter-level deviations from ideal surface shapes such as spheroid or paraboloid.

\begin{figure*}[htbp]
    \centering
    \includegraphics[width=\textwidth]{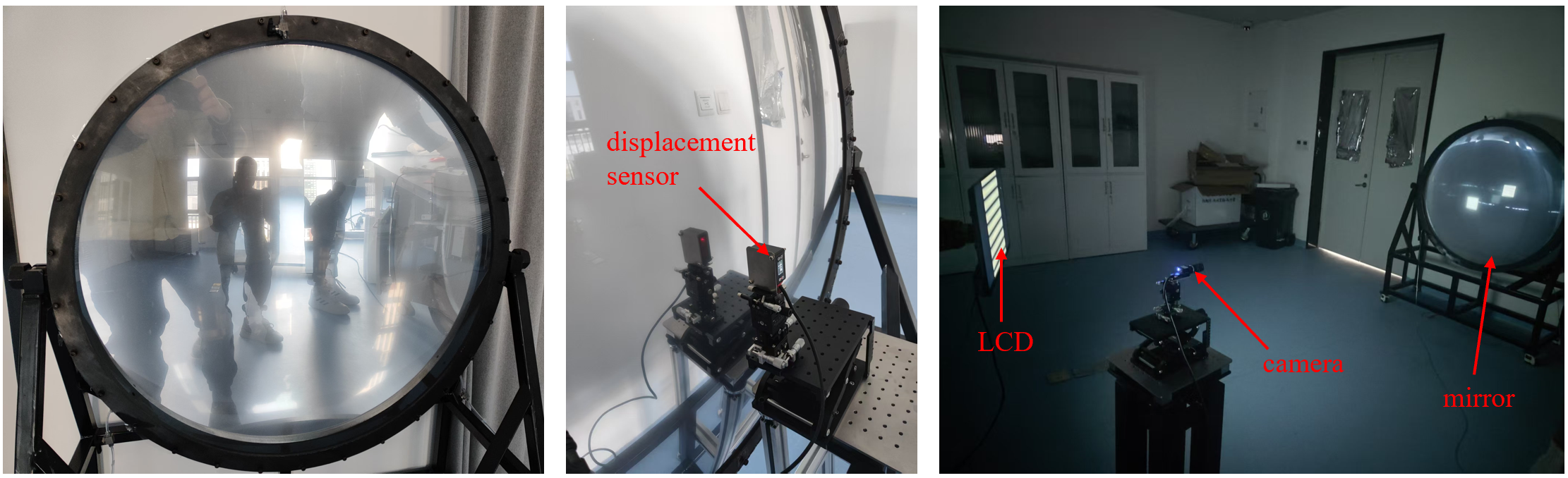}
    {\centering \caption{\label{fig:mirror}
    \textit{Left}: The 1-meter aperture membrane mirror prototype featuring an aluminum-coated PET membrane ($50\ \mu$m thickness) stressed by Nitrogen gas pressure in a Hencky configuration. 
    \textit{Middle}: Rear view showing the high-precision laser displacement sensor ($\pm 1\ \mu$m repeatability) that provides real-time feedback to the PID control system for active surface stabilization. 
    \textit{Right}: Experiment setup comprising an LCD fringe projector and a CMOS industrial camera with a 16~mm F/2.8  lens.
    }}
\end{figure*}

We present a phase-measuring deflectometry (PMD \cite{huang18,zhang25}) solution for a 1-meter ``Hencky-structure'' membrane mirror \cite{hencky15}, where uniform gas pressure provides stress (Fig.~\ref{fig:mirror}). 
PMD is an optical metrology technique evolved from structured light illumination methods, specifically optimized for characterizing the figure profile of freeform specular reflective surfaces.\footnote{In this work, we focus on reflection, but PMD can also measure refractive surfaces \cite{wang21}.}
In general, PMD involves analyzing distorted fringe patterns reflected from the test surface to determine local surface gradients.
The surface figure is then reconstructed from the measured gradients.
This technique uniquely combines non-contact operation (eliminating mechanical distortion risks), freeform surface metrology (accommodating arbitrary surfaces beyond conic sections), and high-temporal-resolution acquisition (second-level measurement cycles allowing stability monitoring).
Therefore, PMD satisfies the requirement of surface figure metrology for membrane mirrors. 

However, traditional PMD methods often suffer from severe ``height-slope ambiguity'' \cite{huang16, fan23},
particularly when characterizing large-size surfaces with steep slopes.
To address this issue, we develop an enhanced PMD framework incorporating a rigorous ray-tracing model and an iterative reconstruction algorithm.
We apply our method to the membrane mirror and measure its surface figure and stability. 

The paper is structured as follows. 
In \S\ref{sec:method}, we describe our methodology and experiment. 
We present our measurement result in \S\ref{sec:res}.
In \S\ref{sec:sum}, we summarize our work and discuss future prospects for membrane mirrors and their surface-figure measurements.

\section{Experiment and Methodology}
\label{sec:method}
\begin{figure}[htbp]
    \centering
    \includegraphics[width=0.5\textwidth]{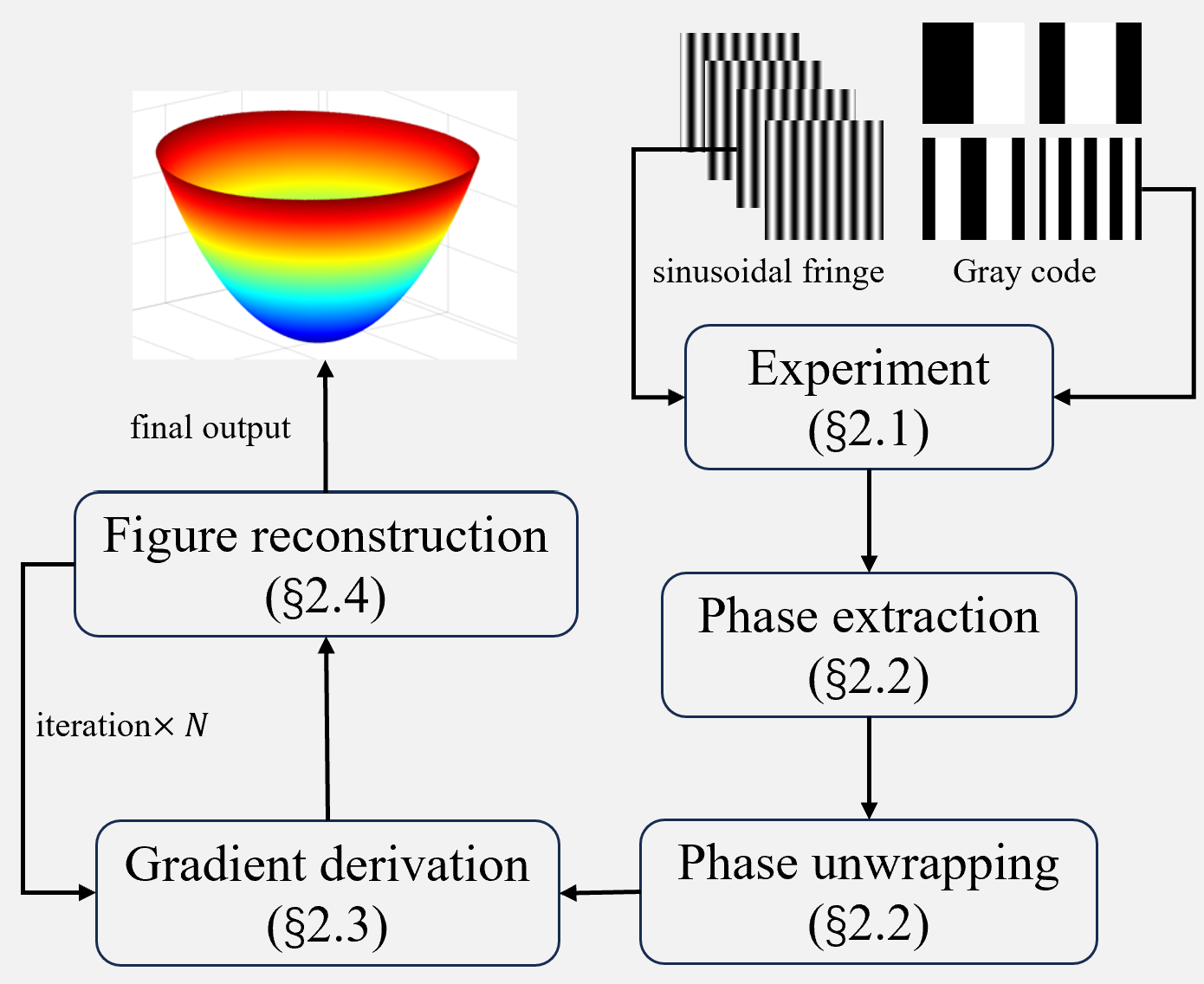}
    {\centering \caption{\label{fig:flow}
    Workflow diagram of our PMD surface reconstruction process for membrane mirrors, showing the sequence from experiment to surface-figure output.}}
\end{figure}

We employ PMD to characterize the surface figure of the membrane mirror. Fig.~\ref{fig:flow} outlines the complete PMD measurement pipeline: 
(1) acquisition of reflected fringe patterns (sinusoidal and Gray code) from the experimental setup (\S\ref{sec:exp}); (2) phase extraction and absolute phase determination through phase unwrapping (\S\ref{sec:phase}); 
(3) surface gradient calculation using our derived phase-to-gradient transfer function (\S\ref{sec:grad}); and (4) final surface reconstruction via an iterative integration algorithm (\S\ref{sec:surface}). Each processing stage is described in detail in the following sections.

\subsection{Experiment setup}
\label{sec:exp}
The test membrane mirror features a 1-meter diameter optical surface formed by two thin polyethylene terephthalate (PET) membranes clamped between three concentric aluminum alloy rings. 
The reflective surface consists of a $50\ \mu$m thick PET substrate coated with a 100~nm aluminum layer, encapsulated by a separate $10\ \mu$m transparent PET membrane to maintain gas pressure. 
The central ring serves dual functions: it secures both membranes and incorporates two access ports---one connected to a nitrogen gas supply via a diaphragm pump and the other to a barometer for pressure monitoring. Both connections use flexible silicone tubing to minimize mechanical coupling.

To compensate for minor gas leakage, the diaphragm pump continuously supplies gas while an automated regulation system maintains the mirror center position within $\approx \pm 2\ \mu$m stability. This closed-loop control system incorporates a high-precision ($\pm 1\ \mu$m repeatability) laser displacement sensor at the mirror backside (Fig.~\ref{fig:mirror} middle) for real-time feedback to 
a proportional-integral-derivative (PID) controller for active surface stabilization.\footnote{\fst{We found the displacement sensor to be significantly more effective than barometric control. This is because constant pressure control could not compensate for the membrane's stress relaxation—a material property change that alters the mirror shape over time. The displacement sensor directly measures the surface central position, allowing the control system to correct for this drift.}}
We align the mirror orientation using a spirit level to ensure perpendicularity to the gravitational vector.
We hereby define a right-handed coordinate system where the mirror center is the origin point, the $x$-direction is horizontal and parallel to the alloy rings, the $y$-direction is vertical, and the $z$-direction is orthogonal to both, pointing outward from the reflective surface.
    
In our experiment, a 32-inch $1366 \times 768$ liquid-crystal display (LCD, pixel pitch $0.51\ $mm) is employed to display fringe patterns (Fig.~\ref{fig:mirror}). 
The display connects to the host computer via High-Definition Multimedia Interface (HDMI). 
\fst{Our fringe spatial frequency was set to 49.8 mm/period, corresponding to eight periods across the LCD screen. This value was chosen to balance two competing factors: excessively high fringe density can introduce phase unwrapping artifacts and discontinuous phase jumps between adjacent pixels; overly sparse fringes result in a reduced phase dynamic range and increased uncertainty in phase retrieval. The selected frequency represents an empirical compromise between these two factors.}
An industrial $2/3''$ $2448 \times 2048$ complementary metal-oxide-semiconductor (CMOS;  pixel pitch $3.45\ \mu$m) camera (Hikrobot MV-CS050-10UC), equipped with a \fst{camera objective with 16~mm focal length}, is used to capture the reflected images. 
The CMOS camera interfaces with the host computer via Universal Serial Bus 3.0 (USB 3.0), providing simultaneous power delivery, data transfer, and camera control.
    
\begin{figure}[htbp]
    \centering
    \includegraphics[width=0.5\textwidth]{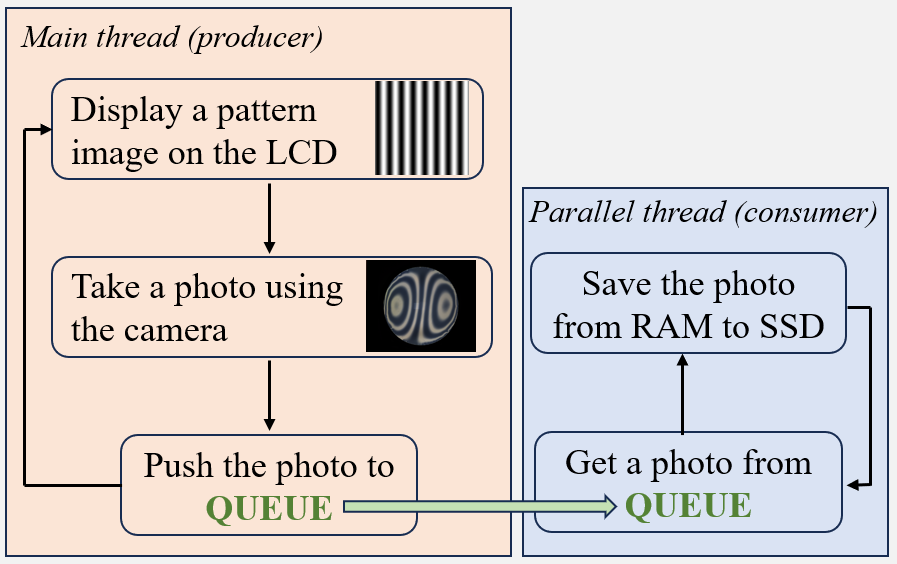}
    {\centering \caption{\label{fig:photo_flow}
    Parallel processing architecture of PMD-SFM for high-throughput PMD measurements. 
    The Python-based system implements a producer-consumer model where: the main thread acquires images from the camera and pushes them into a queue buffer, while a dedicated I/O thread asynchronously writes the queued images from RAM to SSD. 
    This parallel pipeline eliminates storage bottlenecks, significantly reducing total measurement latency.
    }}
\end{figure}

We develop PMD-SFM (Phase-Measuring Deflectometry Surface Figure Monitor), a specialized Python-based software system designed for this study and future PMD experiments. The software integrates all essential PMD measurement functions, including camera/screen control, high-speed data transfer, phase extraction and unwrapping, as well as surface-profile reconstruction.
As illustrated in Fig.~\ref{fig:photo_flow}, PMD-SFM implements an efficient ``producer-consumer'' architecture that enables efficient data transfer from random-access memory (RAM) to solid-state drive (SSD), reducing the total measurement time from $\sim 20$~seconds (sequential processing) to $\sim 8$~seconds per complete cycle.

\subsection{Phase extraction and unwrapping}
\label{sec:phase}
\begin{figure}[htbp]
    \centering
    \includegraphics[width=0.5\textwidth]{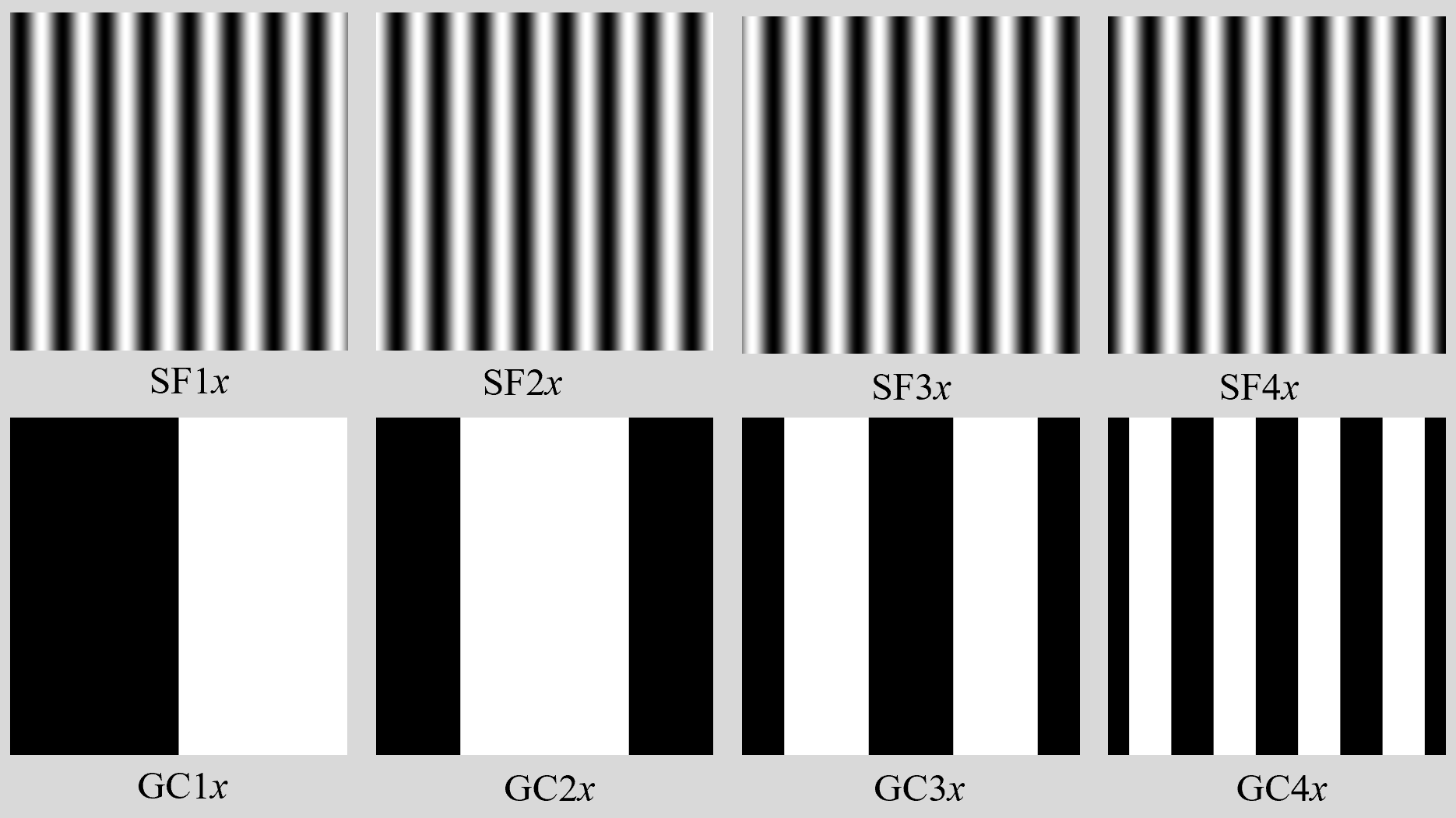}
    {\centering \caption{\label{fig:stripe}
    Structured light patterns displayed on the LCD for absolute phase measurement: (top) Phase-shifted sinusoidal fringes for wrapped phase acquisition and binary Gray codes for phase unwrapping (bottom). 
    Patterns shown are for $x$-direction measurement, with $y$-direction equivalents generated by 90 degree counterclockwise rotation.
    }}
\end{figure}

\begin{figure}[htbp]
    \centering
    \includegraphics[width=0.5\textwidth]{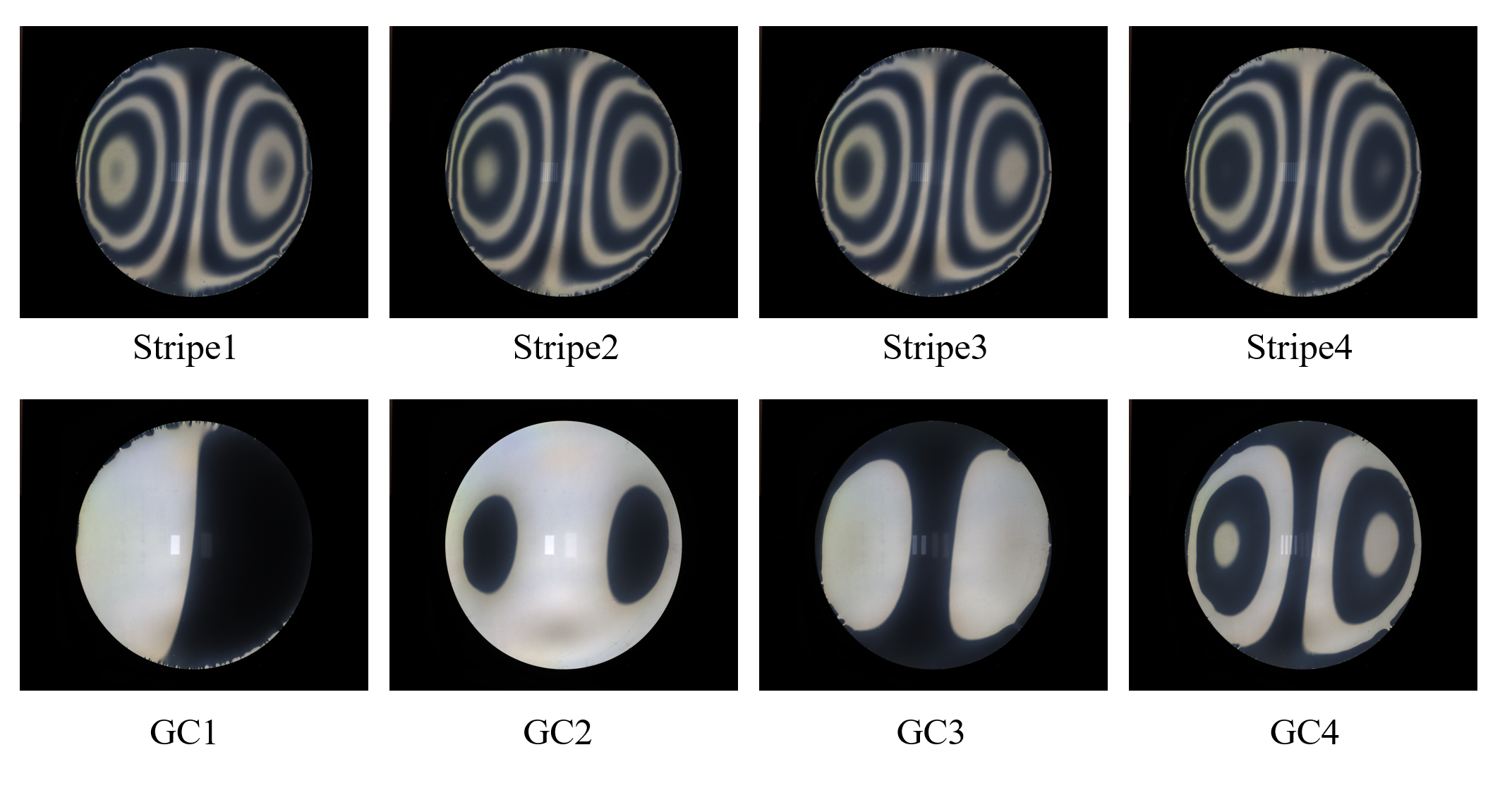}
    {\centering \caption{\label{fig:ccd}
    Camera-captured images of the displayed fringe patterns (see Fig.~\ref{fig:stripe}), showing the reflected sinusoidal-fringe (top) and Gray-code (bottom) patterns used for phase measurement.
    }}
\end{figure}

Our surface figure measurement protocol employs a comprehensive set of 18 pattern images displayed on the LCD and captured by the camera. This set includes four phase-shifted sinusoidal fringe patterns and four Gray-code patterns for both x- and y-direction measurements (Figs.~\ref{fig:stripe} and \ref{fig:ccd}), supplemented by two full-field black and white reference images for intensity calibration.
The sinusoidal fringes enable wrapped-phase extraction and the Gray-code patterns provide absolute phase unwrapping. 
The procedure is detailed below.

\begin{figure*}[htbp]
    \centering
    \includegraphics[width=\textwidth]{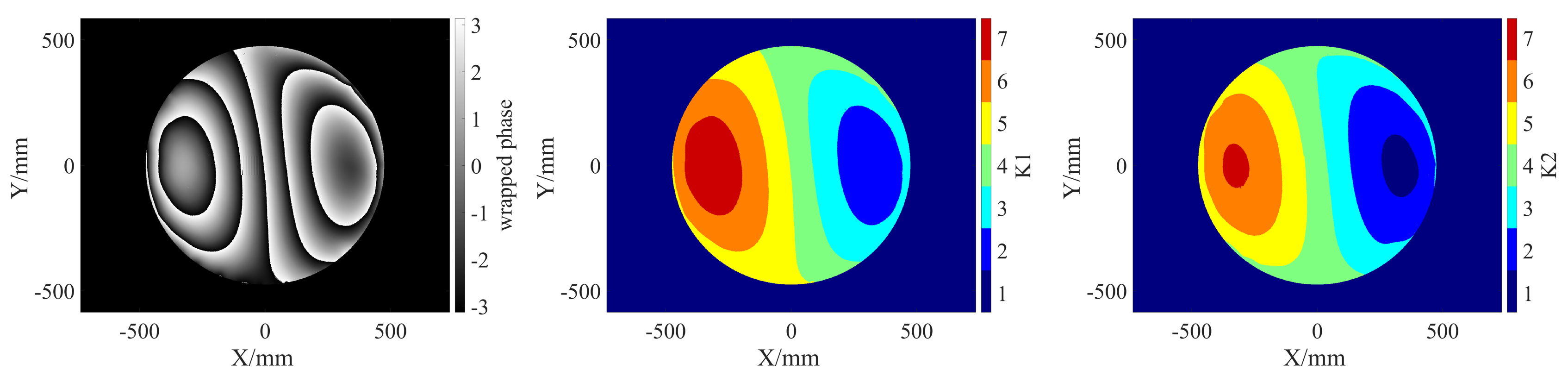}
    {\centering \caption{\label{fig:wrapped_phase}
    Phase calculation components: Wrapped phase $\varphi_{w}(x, y) \in (-\pi, \pi]$ showing periodic oscillations (left);
    the $K_1$ order decoded from GC1-3 patterns (middle);
    the $K_2$ order decoded from GC4 for jump discontinuity resolution (right).
    The combined three components allow us to robustly obtain absolute phase even for regions with large gradients. 
    }}
\end{figure*} 
    
We employ a standard four-step phase-shifting algorithm \cite{wang21} to calculate the wrapped phase $\varphi_{w}(x, y)$ at each measurement point:
\begin{equation}\label{eq:phase_extraction}
\varphi_{w}(x, y)=\arctan \left(\frac{I_{4}(x, y)-I_{2}(x, y)}{I_{1}(x, y)-I_{3}(x, y)}\right),
\end{equation}
where $I_1$ through $I_4$ represent the intensity distributions captured from four phase-stepped sinusoidal fringe patterns with relative phase shifts of 0, $\pi/2$, $\pi$, and $3\pi/2$, respectively. The arctangent operation confines the phase values to the principal value range $(-\pi, \pi]$, creating the characteristic wrapped phase map shown in Fig.~\ref{fig:wrapped_phase}~(left).
This periodic discontinuity requires subsequent phase unwrapping to obtain continuous surface height information.

\begin{figure}[htbp]
    \centering
    \includegraphics[width=0.5\textwidth]{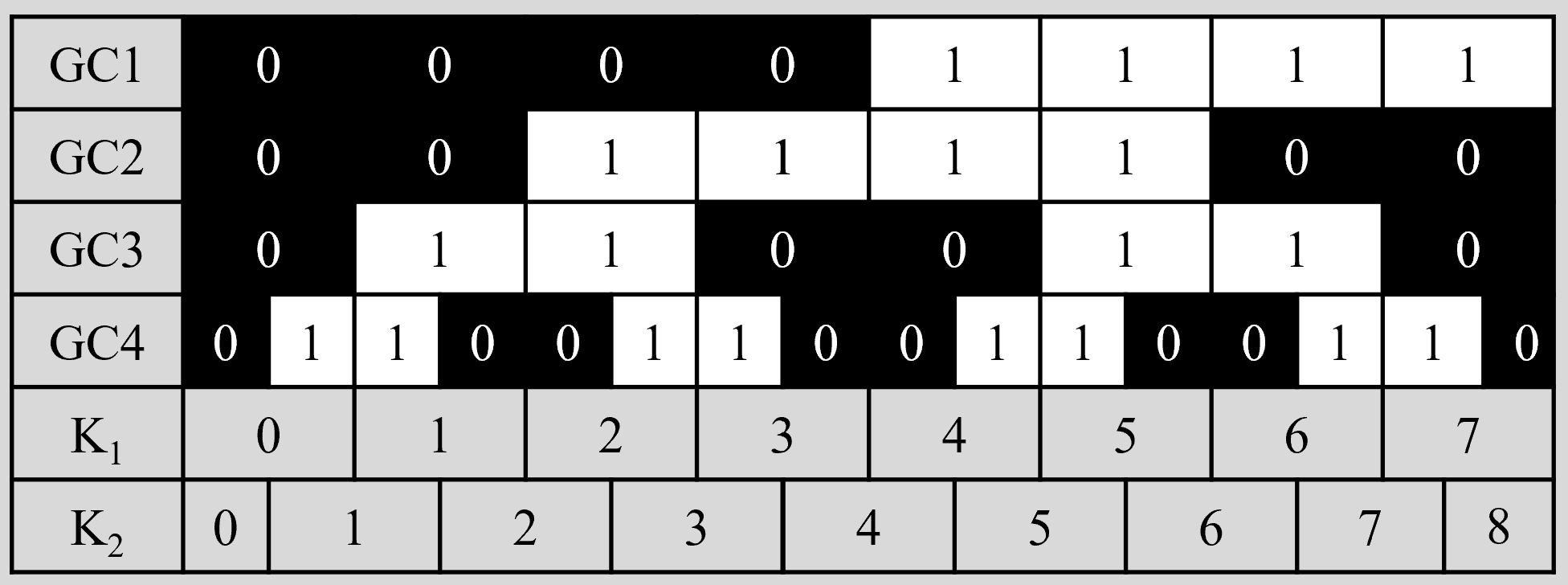}
    {\centering \caption{\label{fig:gc_table}
    Gray code encoding and decoding scheme. The 4-bit Gray-code table establishes absolute phase orders by ensuring single-bit transitions between adjacent values. Decoding the camera-captured Gray code patterns yields the coefficients $K_1$ and $K_2$, which resolve $2\pi$ ambiguities in the wrapped phase map through the unwrapping equation (Eq.~\ref{eq:phase_unwrapping}).
    }}
\end{figure}    
    
We utilize a Gray-code-based phase unwrapping technique \cite{Wu19} to address the challenges of PMD measurements on highly deformable membrane mirrors. As shown in Fig.~\ref{fig:gc_table}, our approach first uses GC1-3 codes to establish the primary fringe order $K_1$ (Fig.~\ref{fig:wrapped_phase} middle), which segments the wrapped phase map (Fig.~\ref{fig:wrapped_phase} left) into continuous regions. However, due to the inherent limitations of modulo-2$\pi$ phase measurements, $K_1$ exhibits significant uncertainty at phase jump discontinuities. To resolve this, we introduce an additional GC4 encoding that specifically targets these jump positions, generating a secondary order $K_2$ (Fig.~\ref{fig:wrapped_phase} right). 
    
\begin{figure}[htbp]
    \centering
    \includegraphics[width=0.5\textwidth]{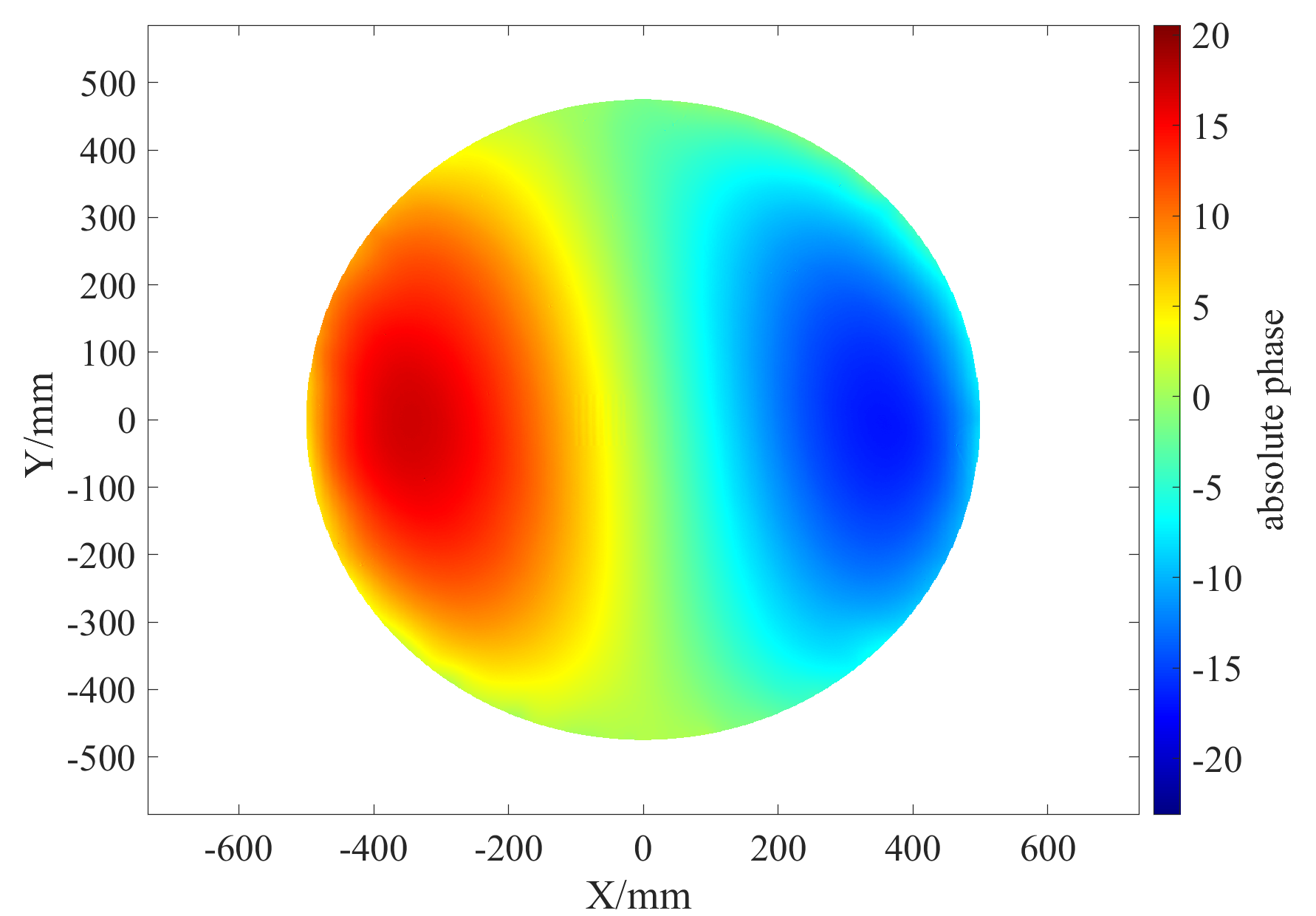}
    {\centering \caption{\label{fig:unwrapped_phase}
    Reconstructed $x$-direction absolute phase map after Gray code-based unwrapping, showing continuous phase distribution across the mirror surface. The color scale represents phase values in radians with discontinuities successfully resolved.
    The phase map for $y$-direction is similar.
    }}
\end{figure}

The complete unwrapping solution combines these components through the equation
    \begin{equation}\label{eq:phase_unwrapping}
        \varphi(x, y) = \varphi_w(x, y) + 2\pi \cdot \left\{\begin{array}{lr} K_{2}(x, y), & \varphi_w(x, y) \leq-\pi / 2 \\ K_{1}(x, y), & -\pi / 2<\varphi_w(x, y)<\pi / 2 \\ K_{2}(x, y)-1, & \varphi_w(x, y) \geq \pi / 2\end{array}\right.
    \end{equation}
\fst{where $\varphi(x, y)$ represents} the absolute (unwrapped) phase. The resulting absolute phase map (Fig.~\ref{fig:unwrapped_phase}) demonstrates the effectiveness of our method, showing smooth, continuous phase distributions even at the mirror's edges where phase gradients are sharp. This robust performance is particularly crucial for membrane mirror characterization, where large surface deformations create challenging measurement conditions that demand advanced unwrapping solutions.
    
\subsection{Gradient derivation}
\label{sec:grad}
In this study, we develop a rigorous theoretical framework to establish the relationship between phase and mirror surface gradients, advancing beyond conventional PMD models that rely on approximations such as weak normal-height dependence and pre-assumed gradient distribution \cite{Knauer04, Tang08}. Our unified approach provides an accurate ray-tracing description, particularly for large-scale deformations. 
Similar as in \S\ref{sec:phase}, we detail the theoretical PMD model below for $x$-direction, while that for $y$-direction follows the same principle.

\begin{figure*}[htbp]
    \centering
    \includegraphics[width=0.9\textwidth]{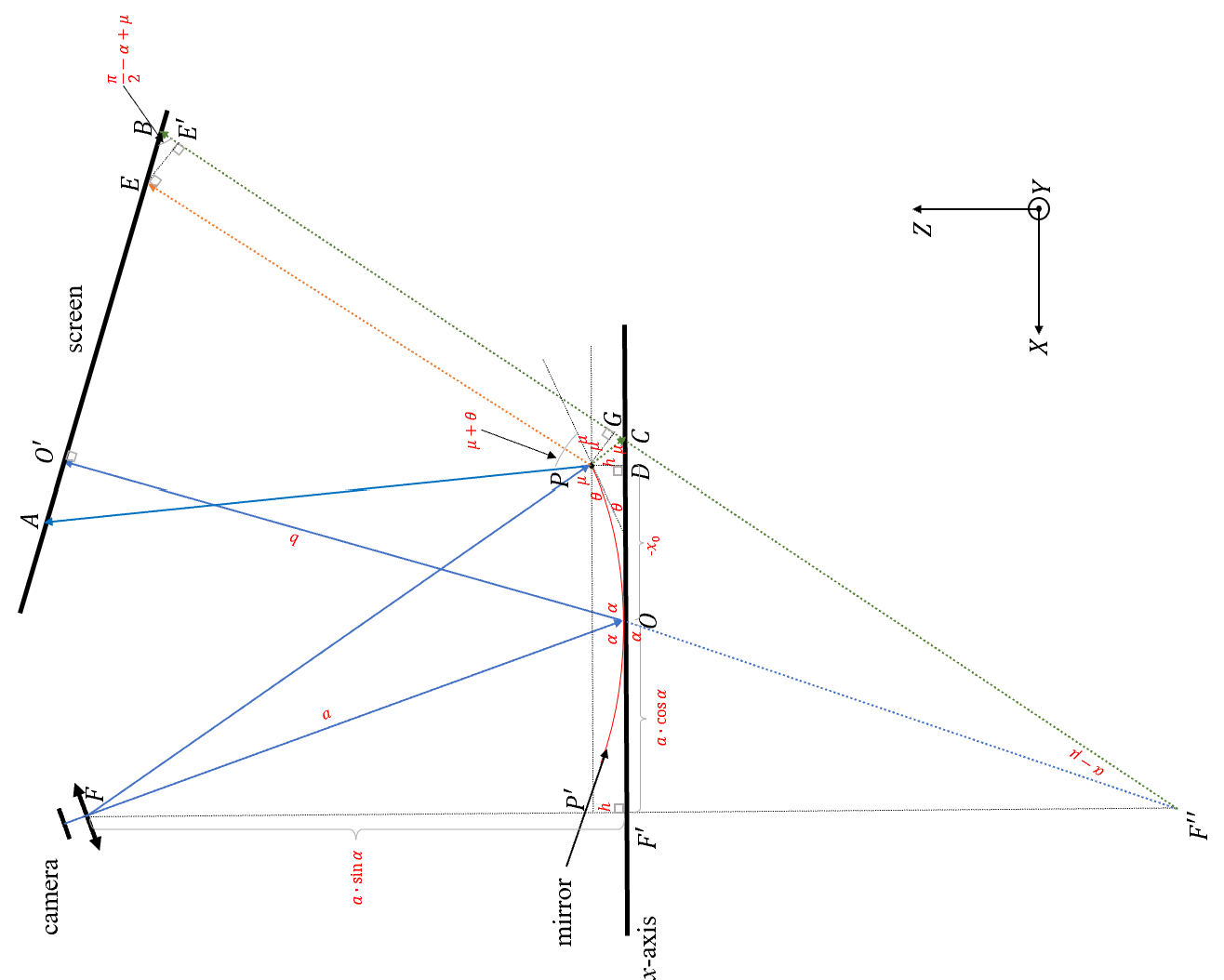}
    {\centering \caption{\label{fig:x}
    Geometric model for phase-to-gradient conversion in $x$-direction. Ray-tracing schematic shows the camera, mirror surface, and display screen projected in the $xz$-plane, with the reflected ray (solid arrow) obeying the law of reflection. The model allows us to rigorously derive the analytical relation between measured phase and surface slope. The model for $y$-direction is similar.
    }}
\end{figure*}

Fig.~\ref{fig:x} displays the geometric configuration.
For convenience, we employ a reverse ray-tracing approach based on the principle of optical path reversibility. Light rays are assumed to originate from the camera pinhole $F$ (using a pinhole model), reflect off the mirror surface, and terminate at the LCD screen. The chief ray is $FOO'$, where $O$ is the mirror vertex (coordinate origin) and $O'$ functions as the phase reference point ($\varphi = 0$).
Key metrological parameters were measured as camera-to-mirror distance $OF\equiv a =2672\pm 2$~mm, screen-to-mirror distance $OO'\equiv b =2871\pm 2$~mm (both measured by a laser distance meter), and angular reference $\angle FOF' \equiv \alpha = 80.9 \pm 0.3 \deg$ (measured by a digital protractor), where the point $F'$ lies on the $x$-axis with $FF'\perp x$-axis. Another auxiliary point $F''$ was constructed via $FF'=F'F''$ extension.

For an arbitrary light ray, the complete optical path can be described as follows: a beam emitted from $F$ reflects at a mirror surface point $P(x_0, h)$ before reaching the screen point $A$.
The displacement $O'A$ is directly proportional to the measured phase $\varphi$ through the relation 
\begin{equation}\label{eq:oa1}
O^{\prime}A=\frac{T}{2\pi}\varphi,
\end{equation}
where $T=49.82$~mm represents the period of the projected fringe pattern on the LCD screen.

This phase shift can be systematically decomposed into three components.
First, the phase deviation caused by the lateral displacement of $P$ from the reference point $O$ is
\begin{equation}
\label{eq:ob}
    O^{\prime}B=(a+b)\tan(\alpha-\mu),
\end{equation}
where 
\begin{equation}
\label{eq:mu}
    \tan \mu = \frac{FP'}{PP'} = \frac{FF'-F'P'}{F'O + OD} = \frac{a\sin \alpha - h}{a\cos\alpha - x_0}
\end{equation}
denotes the angle between the ray $FP$ and the $x$-axis.
Second, the height-induced phase shift is 
\begin{equation}
\label{eq:eb}
\begin{split}
    -EB&= -\frac{EE'}{\sin \angle EBE'}
       = -\frac{GC}{\sin (\pi/2 - \alpha + \mu)}
       = -\frac{PC\sin(2\mu) }{\cos (\alpha - \mu)} \\
       &= -\frac{PD\sin(2\mu)/\sin\mu }{\cos (\alpha - \mu)}
       = -\frac{2h \cos\mu}{\cos(\alpha-\mu)}.
\end{split}
\end{equation}   
Finally, adding up the gradient-dependent component ($-AE$), the observed phase shift is 
\begin{equation}
\label{eq:oa2}
    O'A = O'B - EB - AE.
\end{equation} 
From Eqs.~\ref{eq:oa1}--\ref{eq:oa2}, we can obtain 
\begin{equation}
\label{eq:ae1}
    AE = -O'A + O'B - EB = -\frac{T}{2\pi}\varphi + (a+b)\tan(\alpha-\mu) - \frac{2h \cos\mu}{\cos(\alpha-\mu)},
\end{equation} 
which is directly related to the local surface gradient at point $P$ and serves as the foundation for subsequent gradient calculations.

Assuming the angle between the tangent at point $P$ and $x$-axis is $\theta$, from the law of reflection, the angle of $\angle APE$ is directly related to $\theta$ by
\begin{equation}
\label{eq:ape}
\angle APE = 2\theta. 
\end{equation}
From the geometric relation in Fig.~\ref{fig:x}, we can get another angle in $\triangle APE$ as 
\begin{equation}
\begin{split}
\angle AEP &= \angle O'BF'' = \frac{\pi}{2} - \alpha + \mu
\end{split}
\end{equation}
Knowing $\angle APE$ and $\angle AEP$, we can apply the law of sines to $\triangle APE$, i.e.,
\begin{equation}
\label{eq:ae2}
\begin{split}
AE &= \frac{PE \sin \angle APE}{\sin \angle PAE} = \frac{PE \sin 2\theta}{\sin (\angle APE + \angle AEP) } \\
& = \frac{PE \sin 2\theta}{\sin (2\theta + \pi/2 - \alpha + \mu) } \\
& = \frac{PE \sin 2\theta}{ \sin 2\theta \sin(\alpha-\mu) + \cos 2\theta \cos(\alpha-\mu) } \\
& = \frac{PE \tan 2\theta}{ \tan 2\theta \sin(\alpha-\mu) + \cos(\alpha-\mu) }.
\end{split}
\end{equation}
From the equation above, we can solve for the expression of $\tan \theta$ as
\begin{equation}
\label{eq:tan2theta}
\begin{split}
\tan \theta = \tan\biggl\{ \frac{1}{2} \arctan \biggl[\frac{AE \cos(\alpha - \mu)}{PE-AE \sin(\alpha-\mu)}\biggr] \biggr\}
\end{split}
\end{equation}
where the length $PE$ can be derived as 
\begin{equation}
\label{eq:pe}
\begin{split}
PE &= PG + GE = PG + CE' \\
   &= PC \cos 2\mu + F''B - F''C - E'B \\
   & = \frac{h \cos 2\mu}{\sin \mu} + \frac{F''O'}{\cos (\angle O'F''B)} - FC - \frac{EE'}{\tan\angle EBE'} \\
   & = \frac{h \cos 2\mu}{\sin \mu} + \frac{FO+OO'}{\cos (\angle O'OC - \angle FCO)} \\ 
   & \ \ \ \ - \frac{FF'}{\sin \mu} - \frac{GC}{\tan(\pi/2 - \angle O'F''B)} \\
   & = \frac{h \cos 2\mu}{\sin \mu} + \frac{a + b}{\cos (\alpha - \mu)} - \frac{a \sin \alpha}{\sin \mu} - \frac{PC \sin \angle GPC}{\tan(\pi/2 - \alpha + \mu)} \\
   & = \frac{h \cos 2\mu}{\sin \mu} + \frac{a + b}{\cos (\alpha - \mu)} - \frac{a \sin \alpha}{\sin \mu} - \frac{h \sin 2\mu}{\tan(\pi/2 - \alpha + \mu) \sin \mu}.
\end{split}
\end{equation}
By combining Eqs.~\ref{eq:mu}, \ref{eq:ae1}, and \ref{eq:pe} with Eq.~\ref{eq:tan2theta},
we have the complete function that directly relates the measured phase $\varphi$ to the mirror surface gradient $\tan \theta$.

\fst{
Fig.~\ref{fig:x} and the accompanying analytical expressions describe the optical geometry for reflection points situated within the $xz$-plane. To generalize this formulation to an arbitrary point outside this plane, the configuration in Fig.~\ref{fig:x} can be reinterpreted as lying in a projection plane (denoted as the $p$-plane) that contains the reflection point and is parallel to the $xz$-plane. The only modification required is the introduction of a correction term $\delta\theta$ to account for projection effects, which updates Eq.~\ref{eq:ape} (and consequently Eqs.~\ref{eq:ae2}--\ref{eq:tan2theta}) to the form $\angle APE = 2\theta + \delta\theta$. This angular correction is computed vectorially as}
\begin{equation}
\delta\theta = \arccos(-\vec{n}_p \cdot \vec{u}_p) - \arccos(\vec{n}_p \cdot \vec{v}_p),
\end{equation}
\fst{
where $\vec{n}_p$, $\vec{u}_p$, and $\vec{v}_p$ denote the normalized projections onto the $p$-plane of the surface normal, incident-beam, and reflected-beam vectors, respectively. 
These vectors can be numerically determined from the known surface height distribution $h(x, y)$ (see below).
}

The integrated formulation above provides the mathematical foundation for converting fringe pattern distortions into precise slope measurements of the membrane mirror surface.
\fst{The equation of the local tangent plane is given by $z - x \tan\theta_x - y \tan\theta_y = 0$.}
We note that the height $h(x, y)$ in our equations represents the very surface profile we seek to determine, creating a fundamental ``chicken-and-egg'' problem in the reconstruction process. To break this circular dependency, we implemented an iterative optimization approach that progressively refines the solution through successive cycles of gradient-field estimation and height-domain reconstruction. The procedure is detailed in \S\ref{sec:surface}. 

\subsection{Surface-figure reconstruction}
\label{sec:surface}
For circular membrane mirrors like the one in our experiment, we reconstruct the surface figure using standard Zernike polynomial decomposition, employing the first 36 terms of the series \cite{Wang80}. The surface height profile is expressed as 
\begin{equation}\label{eq:zernike}
h(x, y)=\sum_{k=1}^{36} a_{k} Z_{k}(x, y),
    \end{equation}
where $a_{k}$ represents the coefficient of the $k$-th Zernike polynomial $Z_{k}(x, y)$ defined over the mirror circle. 
By applying partial differentiation to Eq.~\ref{eq:zernike} with respect to both $x$ and $y$ coordinates, we obtain the analytical expressions for surface gradients in orthogonal directions:
\begin{equation}\label{eq:zx}
\tan \theta_{x}=\frac{\partial h(x,y)}{\partial x} =\sum_{k=1}^{36} a_{k} \frac{\partial Z_{k}(x, y)}{\partial x},
\end{equation}

\begin{equation}\label{eq:zy}
\tan \theta_{y}=\frac{\partial h(x,y)}{\partial y} =\sum_{k=1}^{36} a_{k} \frac{\partial Z_{k}(x, y)}{\partial y}.
\end{equation}

The gradient measurement data ($\tan \theta_x$, $\tan \theta_y$) obtained from PMD (\S\ref{sec:grad}) provides the left-hand side of the equations, while the analytical derivatives of Zernike polynomials ($\partial Z_k/\partial x$, $\partial Z_k/\partial y$) are known quantities. 
This formulation leaves the 36 Zernike coefficients ($a_k$) as the only unknowns. For our experimental setup with $N \approx 2.4$ million measurement points this yields an overdetermined system of $2N$ linear equations, which we solve through matrix-based least-squares optimization below.

By organizing the phase-gradient relationships (Eqs.~\ref{eq:zx} and \ref{eq:zy}) for all measurement points, we obtain the linear system:
\begin{equation}\label{eq:matrix}
\underbrace{
\begin{bmatrix}
\frac{\partial Z_1(x_1,y_1)}{\partial x} & \cdots & \frac{\partial Z_{36}(x_1,y_1)}{\partial x} \\
\vdots & \ddots & \vdots \\
\frac{\partial Z_1(x_N,y_N)}{\partial x} & \cdots & \frac{\partial Z_{36}(x_N,y_N)}{\partial x} \\
\frac{\partial Z_1(x_1,y_1)}{\partial y} & \cdots & \frac{\partial Z_{36}(x_1,y_1)}{\partial y} \\
\vdots & \ddots & \vdots \\
\frac{\partial Z_1(x_N,y_N)}{\partial y} & \cdots & \frac{\partial Z_{36}(x_N,y_N)}{\partial y}
\end{bmatrix}
}_{\mathbf{D}\ \in\ \mathbb{R}^{2N \times 36}}
\underbrace{
\begin{bmatrix}
a_1 \\
\vdots \\
a_{36}
\end{bmatrix}
}_{\mathbf{A}\ \in\ \mathbb{R}^{36 \times 1}}
=
\underbrace{
\begin{bmatrix}
\frac{\partial h(x_1,y_1)}{\partial x} \\
\vdots \\
\frac{\partial h(x_N,y_N)}{\partial x} \\
\frac{\partial h(x_1,y_1)}{\partial y} \\
\vdots \\
\frac{\partial h(x_N,y_N)}{\partial y}
\end{bmatrix}
}_{\mathbf{G}\ \in\ \mathbb{R}^{2N \times 1}},
\end{equation}
where $\mathbf{D}$ is the $2N \times 36$ design matrix containing Zernike derivative terms, $\mathbf{A}$ is the $36 \times 1$ coefficient vector, and $\mathbf{G}$ is the $2N \times 1$ gradient measurement vector. The optimal coefficients are obtained via the normal equation:
\begin{equation}\label{eq:matrix2}
\mathbf{A}=\left(\mathbf{D}^{\intercal} \mathbf{D}\right)^{-1}\left(\mathbf{D}^{\intercal} \mathbf{G}\right),
\end{equation}  
This solution provides the complete set of Zernike coefficients for surface reconstruction using Eq.\ref{eq:zernike}. The matrix formulation ensures computational efficiency  on a modern workstation, e.g., the computing time for our case is only $\sim 10$ seconds.

\begin{figure}[htbp]
\centering
\includegraphics[width=0.5\textwidth]{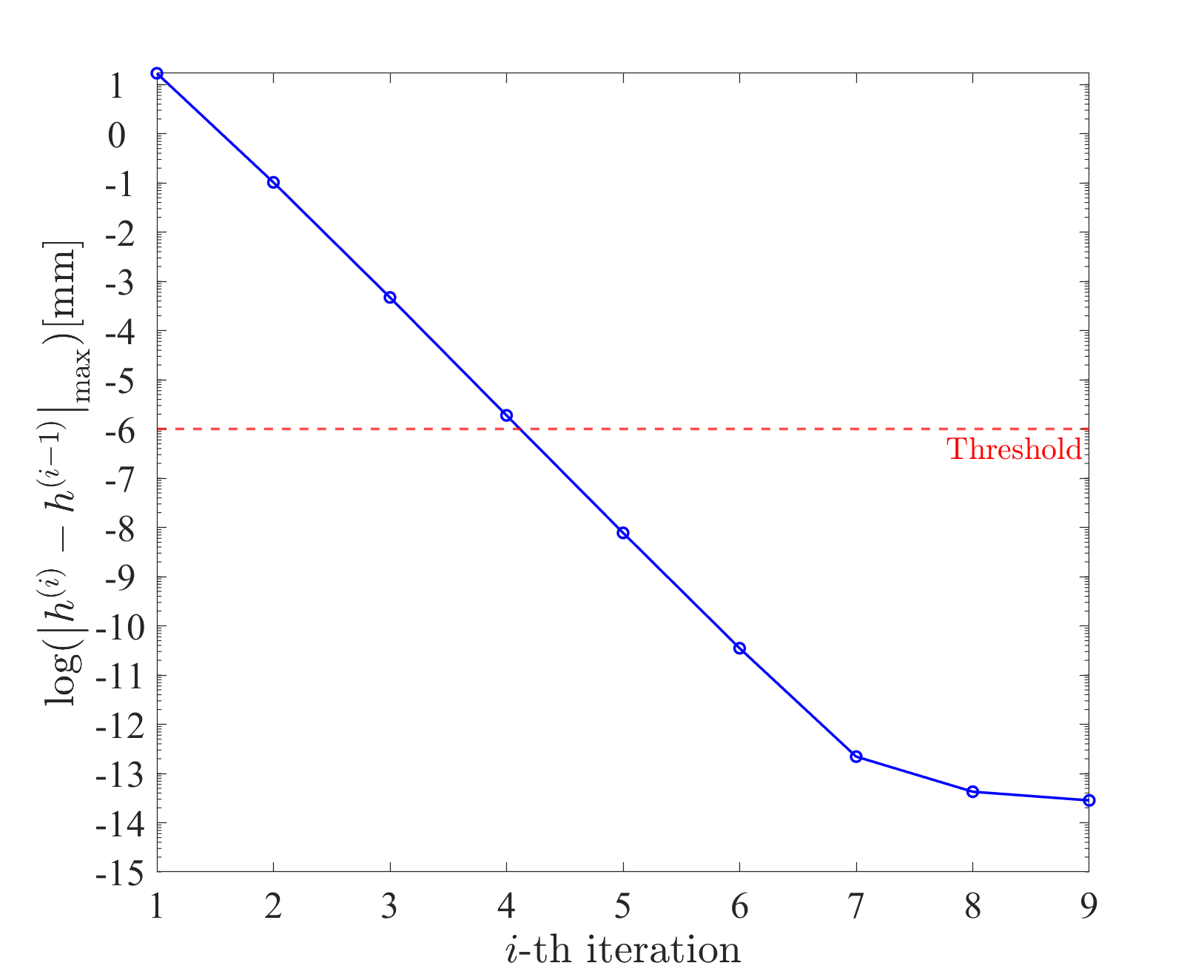}
{\centering \caption{
\label{fig:Iterative2}
Iterative convergence performance showing the maximum absolute surface height difference between consecutive iterations. The exponential decrease in residual error (blue curve) demonstrates rapid convergence during early iterations (1--7), transitioning to asymptotic behavior beyond the 7th iteration. Our convergence threshold ($\epsilon < 10^{-6}$~mm, red dashed line) is satisfied after just 5 iterations, confirming the efficiency of our reconstruction algorithm. 
}} 
\end{figure}

The surface reconstruction faces a fundamental challenge: the gradient calculation in \S\ref{sec:grad} requires prior knowledge of the surface height $h$, which is exactly the unknown quantity we seek to determine. To resolve this circular dependency, we implement an iterative reconstruction scheme initialized with a parabolic approximation 
\begin{equation}
h^{(0)} = \frac{x^2+y^2}{R^2/h_{\max}^{(0)}}
\end{equation}
where $R=500$~mm is the radius of the mirror and we adopt $h_{\max}^{(0)}=40$~mm as the maximum height roughly estimated by ruler measurements. Each iteration cycle consists of three steps: (1) gradient computation using the current height estimate via the geometric model in \S\ref{sec:grad}, (2) solving Eq.~\ref{eq:matrix2} for updated Zernike coefficients $a_k^{(i)}$, and (3) reconstructing an improved surface
\begin{equation}
\label{eq:hi1}
h^{(i+1)}(x,y) = \sum_{k=1}^{36} a_k^{(i)} Z_k(x,y)
\end{equation}
Convergence is achieved when the maximum height variation between iterations satisfies 
\begin{equation}\label{eq:limit}
\left | h^{(i)}-h^{(i-1)} \right |_{\max} < \epsilon
\end{equation} 
where we adopt a sufficiently low threshold of $\epsilon=10^{-6}$~mm. 
This threshold typically reached within 5 iterations, as evidenced by the fast exponential drop in Fig.~\ref{fig:Iterative2}.

\begin{figure}[htbp]
\centering
\includegraphics[width=\textwidth]{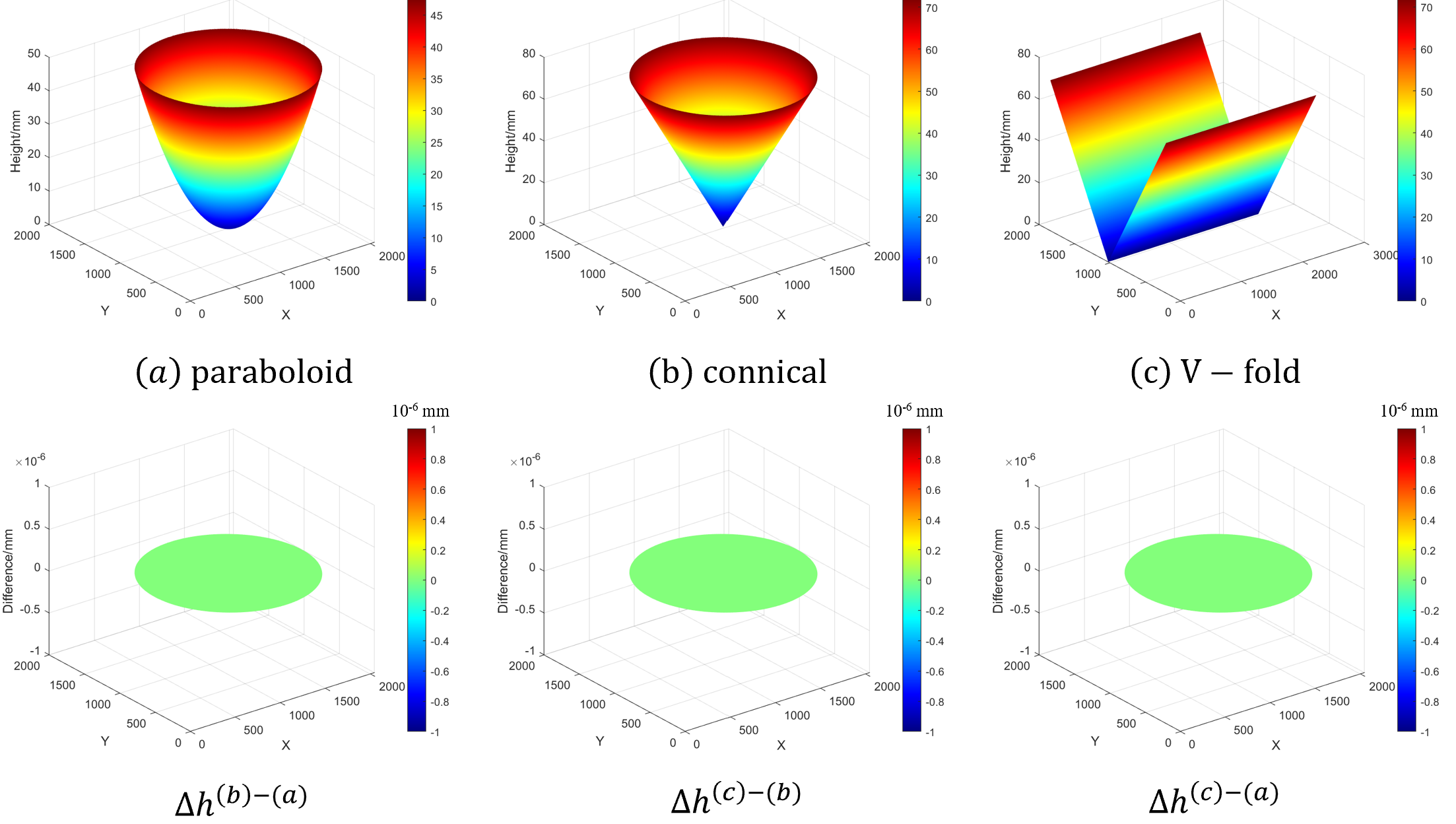}
{\centering \caption{
\label{fig:diff_init_shape}
\fst{\textit{Top}: Three assumed initial surface shapes of paraboloid, conical, and V-fold. \textit{Bottom}: The converged height differences between the reconstructed surfaces after nine iterations. The differences are numerically negligible ($\lesssim 10^{-16}$~mm), confirming convergence to an identical solution regardless of the initial shape. This demonstrates that our reconstruction algorithm is robust and independent of the initial shape assumption.}
}} 
\end{figure}

While the surface reconstruction algorithm above initially adopts a parabolic profile with a maximum height of $h_{\max}^{(0)}=40$~mm as the starting guess, we rigorously verify its robustness by testing various initial conditions: (1) parabolic profiles with maximum heights ranging from 10--100~mm, and (2) fundamentally different surface shapes including conical and V-fold profiles. Remarkably, all test cases consistently converged to identical solutions \fst{(see Fig.~\ref{fig:diff_init_shape} as an example)}.
The consistent convergence behavior demonstrates the algorithm's exceptional insensitivity to initial conditions, a critical advantage for membrane mirror measurements where a priori knowledge of the surface shape is often limited. 
    
\section{Result}
\label{sec:res}
Our PMD methodology (\S\ref{sec:method}) enables characterization of the membrane mirror's surface figure. 
In this section, we evaluate both static and dynamic characteristics of the membrane mirror through two complementary analyses: surface profile with associated uncertainty quantification via Monte Carlo simulations (\S\ref{sec:resfig}), and temporal stability assessment through consecutive PMD measurements with corresponding error analysis (\S\ref{sec:resstb}). 

\subsection{Surface figure}
\label{sec:resfig}

\begin{figure*}[htbp]
\centering
\includegraphics[width=\textwidth]{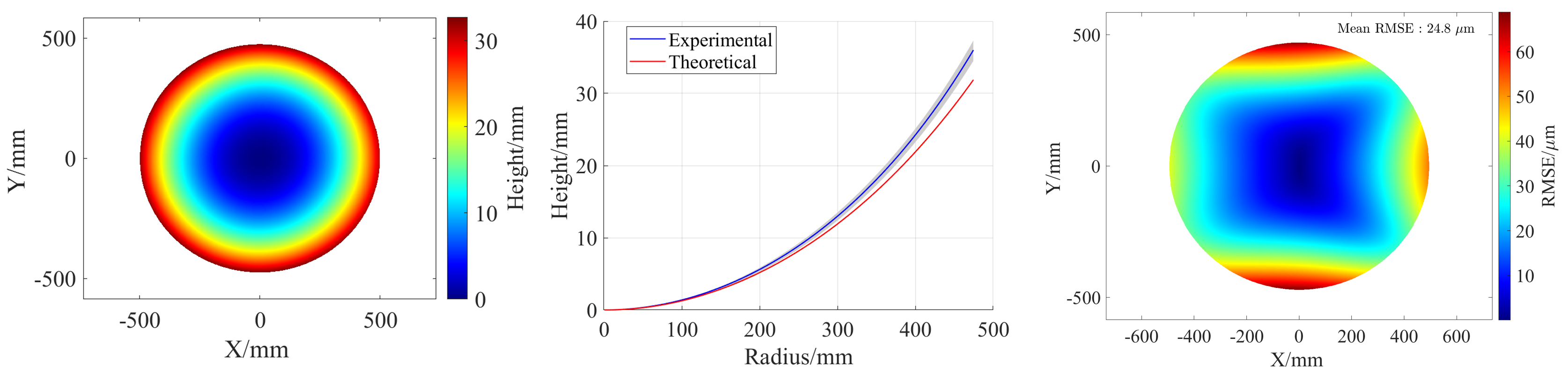}
{\centering \caption{\label{fig:FSS}
Surface characterization results. 
\textit{Left}: Our measured surface height distribution (sag map) of the membrane mirror, with color scale indicating height magnitude. Edge regions beyond 475 mm radius were excluded from our analysis due to the significant noise induced by membrane wrinkles.
\textit{Middle}: Radial profile comparison: the blue curve shows the azimuthally averaged height value as a function of radius with shaded region representing $\pm 1\sigma$ variation, while the red curve displays the theoretical Hencky membrane solution, assuming an elastic modulus of $E = 4.1$~GPa (measured by a universal testing machine) and a Poisson's ratio of $\nu = 0.3$ under pressure of $P=600$~Pa (measured by the barometer). 
\textit{Right}: Monte Carlo-derived Distribution RMSE for the surface height in the left panel. 
Different color indicates different error magnitude. 
The mean RMSE is 24.8~$\mu$m.
The mirror center has zero error by definition, because we take it as zero height reference.
}}
\end{figure*}

Fig.~\ref{fig:FSS} displays (left) the reconstructed two-dimensional surface height distribution. Through azimuthal averaging of the measured surface profile (Fig.~\ref{fig:FSS} middle), we observe a $\approx 3$~mm ($\approx 10$\%) height discrepancy at the mirror edge relative to the ideal Hencky solution \cite{hencky15, fichter97}, despite both profiles sharing a common zero reference at the mirror center. This deviation primarily originates from practical manual mounting of the mirror to the aluminum alloy
rings that differ from theoretical assumptions, i.e., the membrane acquires initial concavity due to self-gravity and  manual handling introduces unavoidable pre-tension. These assembly-induced effects likely account for the observed height differences, particularly in the peripheral regions where the surface gradients are steepest. This comparison highlights the importance of accounting for real-world installation errors in membrane mirror design and modeling.

Due to the complexity of our surface reconstruction algorithm involving iterative calculations and nonlinear parameter dependencies (\S\ref{sec:method}), we implement a Monte Carlo approach for comprehensive uncertainty quantification. The simulation protocol involves three key steps: : (1) random sampling of geometric parameters (i.e., $a$, $b$, and $\alpha$; \S\ref{sec:grad}) from Gaussian distributions centered on measured values with dispersions reflecting their respective measurement uncertainties; (2) incorporation of imaging system noise through pixel-level Gaussian perturbations (adopting an empirical dispersion of  $\sigma=0.4$~ADU, analog-to-digital unit); and (3) 100 independent reconstruction trials to statistically characterize error propagation. We quantify the measurement root-mean-square error (RMSE) by calculating the standard-deviation value of the height distribution for each point obtained from the ensemble of Monte Carlo-sampled surface profiles.

The result (Fig.~\ref{fig:FSS} right) reveals an average RMS error of $\approx 25\ \mu$m across the mirror surface, with mirror center (i.e., the height reference) maintaining zero error by definition. The maximum errors of $\approx 50\ \mu$m are localized in peripheral regions - values substantially smaller than the observed $\approx 3$~mm discrepancies between measured and theoretical profiles. This significant separation between Monte-Carlo uncertainties and systematic deviations confirms that the profile differences primarily reflect physical installation effects (as discussed above) rather than measurement limitations. 

\subsection{Stability}
\label{sec:resstb}

Considering the membrane mirror's sensitivity to environmental perturbations such as airflow disturbances and acoustic vibrations, we implement a specialized stability assessment protocol using our PMD-SFM system. The automated measurement procedure cyclically displays sinusoidal fringe patterns (Fig.~\ref{fig:stripe}) and acquire the reflected photos over a continuous 30-minute period, resulting in 451 image sets (i.e., each measurement only takes $\approx 4$~seconds). In this procedure, the Gray-code patterns and the full-field black/white references are employed only during the initial cycle - a validated approach that does not affect decoding accuracy but improve the measurement speed by about a factor of two. 

\begin{figure*}[htbp]
\centering
\includegraphics[width=\textwidth]{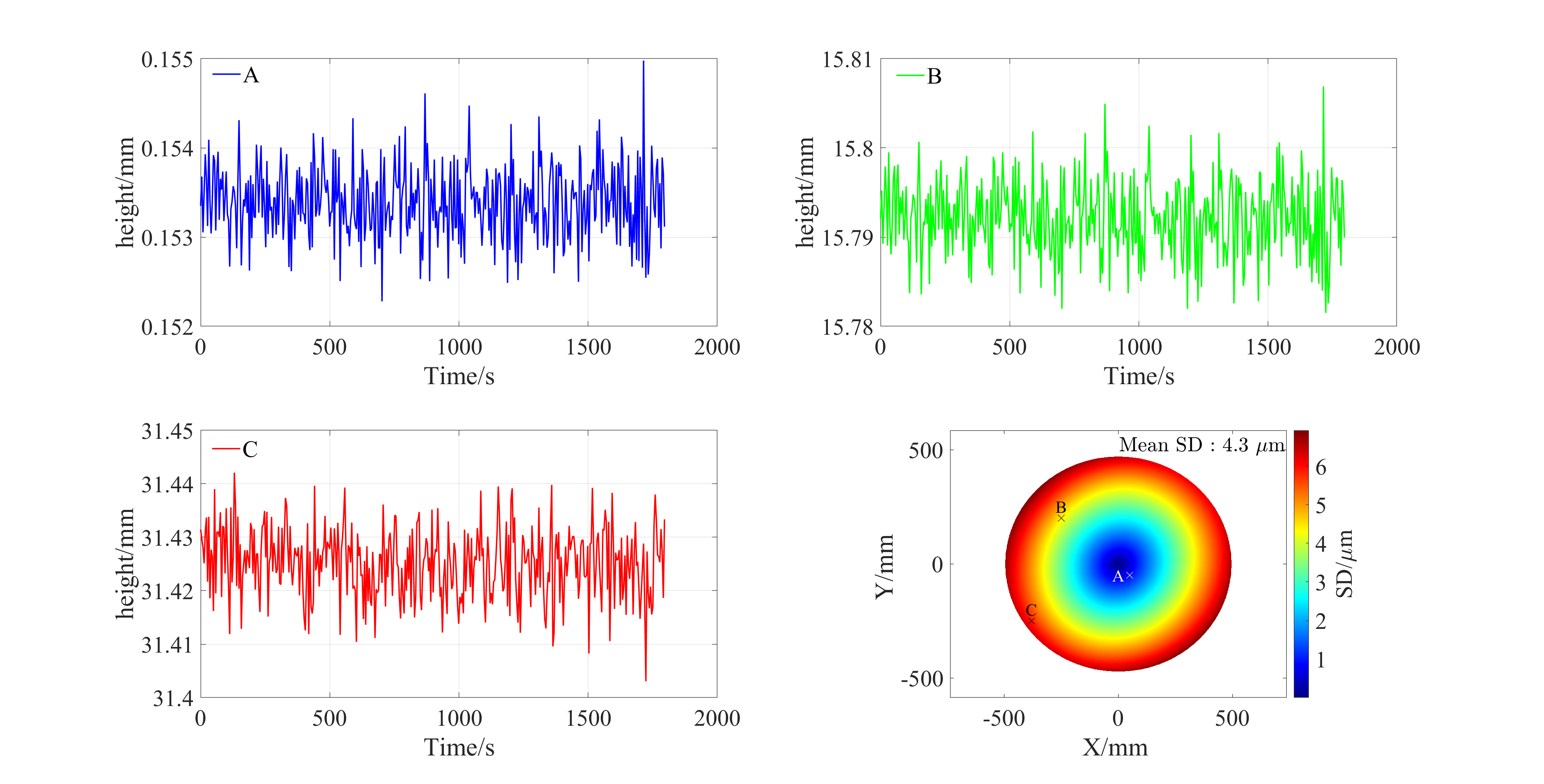}
{\centering \caption{\label{fig:surface_height}Dynamic stability characterization. (Top and bottom-left panels) Temporal evolution of surface height variations at three representative positions (A, B, and C) on the membrane mirror. (Bottom-right panel) Spatial map of standard deviation across the mirror surface, with color scale quantifying local fluctuation magnitudes. Markers indicate the monitored positions A--C. The central region is in general more stable than the peripheral areas, because the mirror center is defined as zero height reference in our analysis. 
}}
\end{figure*}  

\begin{figure*}[htbp]
\centering
\includegraphics[width=0.8\textwidth]{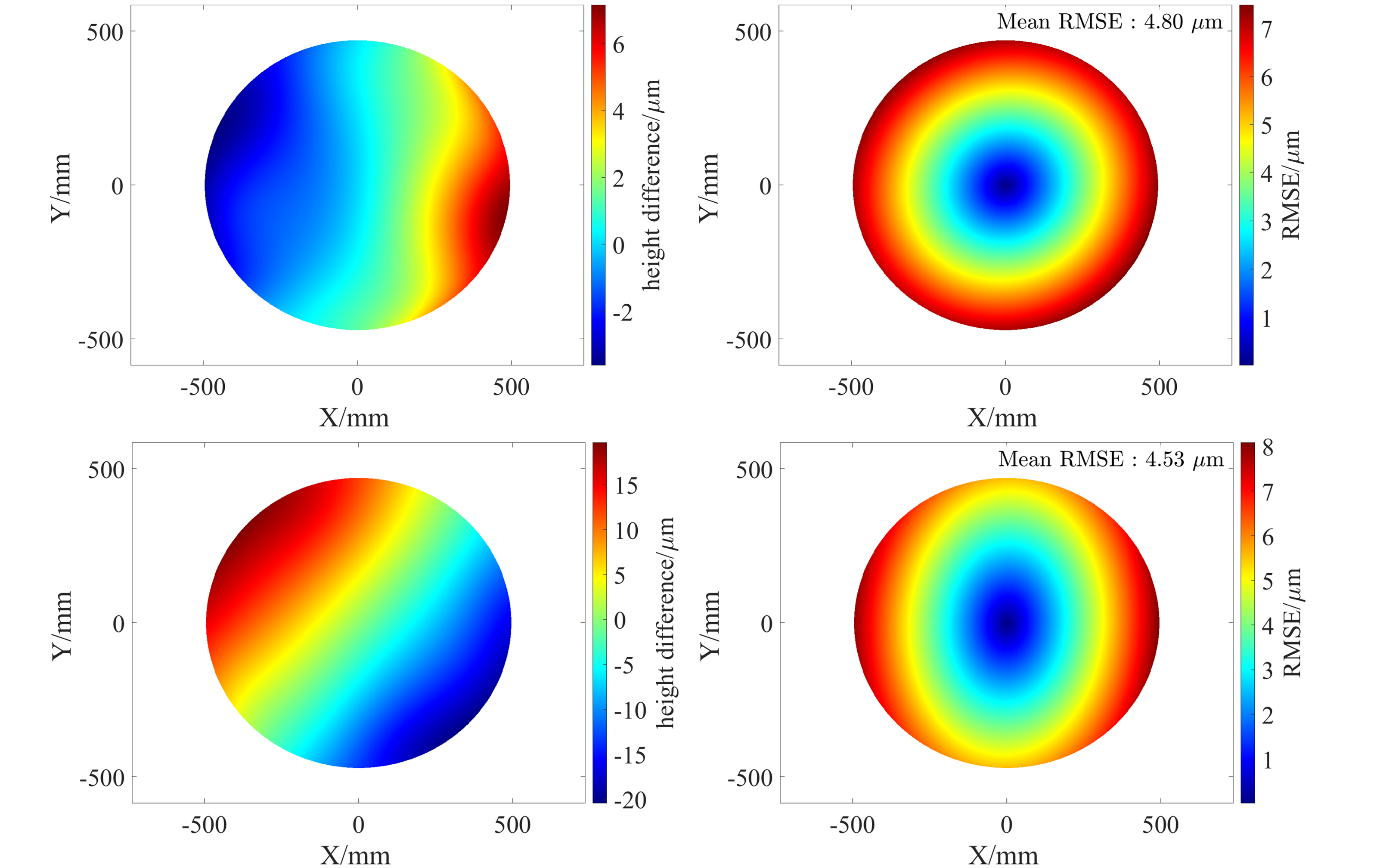}
{\centering \caption{\label{fig:bootstrap2}
\textit{Left}: Surface height changes between two randomly selected measurement pairs with time intervals of 1 minute (top) and 10 minutes (bottom). \textit{Right}: Corresponding uncertainty (RMSE) distributions obtained from Monte Carlo simulations, quantifying the uncertainty in measured surface variations. Both cases yield an average RMSE of $\approx 4$--$5\ \mu$m, indicating consistent measurement precision across different time scales and confirming that observed fluctuations are dominated by system uncertainty rather than intrinsic mirror instability.
}}
\end{figure*}   

Fig.~\ref{fig:surface_height} presents the membrane mirror's stability performance.
The standard deviation (SD) distribution map (bottom-right panel) quantifies the spatial variation in stability. The mirror center, serving as the zero-height reference, shows null SD by definition, with values progressively increasing toward the periphery (reaching $\gtrsim 6\ \mu$m at edges). The average SD across the mirror is $4.3\ \mu$m, reflecting a combination of intrinsic membrane vibrations and measurement system uncertainties.
The temporal fluctuations of surface height at three representative positions are shown in the top and bottom-left panels of Fig.~\ref{fig:surface_height}. The variations appear to be dominated by random noise rather than smooth, systematic changes, suggesting a strong influence of measurement uncertainty. To verify this hypothesis and quantitatively distinguish intrinsic mirror dynamics from measurement noise, we conduct a detailed uncertainty analysis below.

Similar to the approach in \S\ref{sec:resfig}, we employ Monte Carlo simulations to estimate the uncertainties in the stability measurements. However, due to the extreme computational demands associated with simulating all $>400$ PMD measurement sets, we restrict the analysis to 2 randomly selected measurement pairs to evaluate typical errors in surface change detection. As shown in Fig.~\ref{fig:bootstrap2}, the left panels present the reconstructed surface changes for two representative time intervals (1 and 10 minutes), while the right panels display the corresponding Monte Carlo-derived error distributions. The average RMS errors for differential surface measurements are $\approx 4$--$5\ \mu$m, significantly smaller than the absolute surface height error ($\approx 25\ \mu$m; \S\ref{sec:resfig}). This reduction occurs because uncertainties in geometric parameters are largely canceled when computing shape differences between two measurements. 

This error level is comparable to the observed temporal fluctuation ($\rm SD = 4.3\ \mu$m), indicating that the mirror's intrinsic stability is beyond the current resolution limit of our PMD system. This result confirms the membrane mirror maintains $\mu$m-level stability under laboratory conditions, though further improvements in measurement precision (especially the imaging accuracy) are needed to fully resolve its dynamic behavior.
\fst{For astronomical applications, where surface accuracy is typically required to be better than $\lambda/30$, our measurement uncertainties indicate that the current PMD system can meet the precision requirements for far-infrared and longer wavelengths.}

\section{Summary and Future Prospects}
\label{sec:sum}
In this work, we develop a PMD-based method for efficiently measuring the surface profile of membrane mirrors. This approach offers three key advantages: it is non-contact, has a large dynamic range, and operates at high speed. A summary of our methodology and key findings is provided below.

\begin{itemize}

\item Our methodology employs a standard four-step phase-shifting technique to extract the wrapped phase distribution from captured fringe patterns (\S\ref{sec:phase}). We subsequently apply a Gray-code-based unwrapping algorithm to resolve phase ambiguities and obtain absolute phase values. Using a rigorously derived analytical model, we then transform the absolute phase data into surface gradient maps (\S\ref{sec:grad}). The surface profile is ultimately reconstructed through a robust iterative fitting procedure that is independent of the assumption of initial surface profile (\S\ref{sec:surface}).

\item We quantitatively compare the reconstructed surface profile with the theoretical Hencky membrane solution (\S\ref{sec:resfig}). The maximum discrepancy of approximately 3 mm occurs in peripheral regions, significantly exceeding our estimated measurement uncertainty of $\sim 50\ \mu$m derived through comprehensive Monte Carlo simulations. These simulations systematically incorporate uncertainties from measured geometric parameters as well as image pixel fluctuations, confirming that the observed deviations represent physically meaningful discrepancies rather than measurement artifacts. We attribute this discrepancy primarily to assembly-related factors during the installation process, including non-uniform pre-tension and gravity-induced initial deformation.

\item We evaluate the mirror's dynamic stability by performing continuous PMD measurements over a 30-minute period, with each full-surface acquisition completed in approximately 4 seconds (\S\ref{sec:resstb}). The associated uncertainty in surface shape variation is quantified using Monte Carlo simulations.
The results reveal a measurement uncertainty of approximately $\approx 4$--$5\ \mu$m, which aligns closely with the observed temporal variation ($\approx 4\ \mu$m). This similarity indicates that the mirror's intrinsic stability operates at the micrometer level, beyond the current resolution limit of our PMD system. To accurately resolve the mirror's true dynamic behavior, future improvements in measurement precision will be essential.

\end{itemize}

While our current PMD method achieves precise surface measurements, its practical implementation faces two main limitations. First, it requires meticulous manual calibration of geometric parameters ($a$, $b$, and $\alpha$), which is time-consuming and impractical for observatory deployment; and computational latency, with surface reconstruction requiring several minutes per measurement—significantly longer than the 4-second data acquisition time in stability assessment.

To address these challenges, we plan to develop a self-calibrated PMD framework that automatically determines geometric parameters during measurement, eliminating the need for error-prone manual setup \cite{wang21}. Furthermore, we will integrate machine learning-based reconstruction algorithms to accelerate data processing, enabling real-time surface figure computation and closing the gap between acquisition and analysis \fst{\cite{qiao20, nguyen23, zhang24}}. These improvements will enhance the method's efficiency and practicality for long-term, automated operation in astronomical observatories.

\begin{backmatter}
\bmsection{Funding} 
National Natural Science Foundation of China (NSFC), Chinese Academy of Sciences (CAS).

\bmsection{Acknowledgment}
YX, ZZK, and YG acknowledges support from the National Natural Science Foundation of China (NSFC).
We thank Nan-Long Sun and Xiang-Chao Zhang for helpful discussion. 
We thank Nanjing Congren Photoelectric Technology Co., Ltd.\ for their help in design and manufacture of the membrane mirror.

\bmsection{Disclosures}
The authors declare no conflicts of interest.

\bmsection{Data Availability}
Data underlying the results presented in this paper are not publicly available at
this time but may be obtained from the authors upon reasonable request.

\end{backmatter}


\bibliography{ref}

\end{document}